
\hsize27pc
\vsize37pc
\magnification=1200
\bigskip
\centerline{\bf Hydrodynamics and Nonlocal Conductivities in Vortex States}
\centerline{\bf of Type II Superconductors}
\bigskip
\centerline{\rm Ryusuke Ikeda}
\smallskip
\centerline{\rm Department of Physics, Kyoto University, Kyoto 606-01}
\bigskip
\centerline{(\rm Received )}
\bigskip
\bigskip
\bigskip
\leftline{\rm ABSTRACT}
A hydrodynamical description for vortex states in type II superconductors is
presented based on the time-dependent Ginzburg-Landau equation (TDGL).
In contrast to the familiar extension of a single vortex dynamics based
on the force balance, our description is consistent with the known
hydrodynamics of a rotating neutral superfluid and correctly includes
informations on the Goldstone mode. Further it enables one to examine
nonlocal conductivities perpendicular to the field in terms of Kubo formula.
 Typically, the nonlocal conductivities deviate from the usual vortex flow
expressions,  as the nonlocality parallel to the field becomes weaker
than the perpendicular one measuring a degree of positional correlations,
and, for instance, the superconducting contribution of dc Hall conductivity
nonlocal only in directions perpendicular to the field becomes vanishingly
small  in the situations with large shear viscosity, leading to an
experimentally measurable relation $\rho_{xy} \sim{\rho_{xx}^2}$ among
the resistivity components.  Other situations are also discussed on the
basis of the resulting expressions.
\bigskip
\leftline{KEYWORDS: type II superconductor, vortex states, vortex flow
conductivities}
\bigskip
\vfill\eject

\leftline{\bf \S 1. Introduction}

At present, understanding theoretically the nonlocality$^{1,2)}$ of linear
resistivity
in a clean system (with no pinning) seems to be one of central problems on the
vortex
states of a type II superconductor. By analogy to the usual viscous fluid,
Marchetti
and Nelson$^{1)}$ have previously argued within a linear hydrodynamics that,
deep in
the liquid regime, a shear-viscous force proportional to $-\partial^2_{\perp}
\partial_t{\bf s}$ will take the place of the shear term in the elastic force
${\bf f}_{\rm el}$ in the following force-balance equation for the vortex
lattice
$$\partial_t(-\Gamma_1\,{\bf s}+\Gamma_2\,(\hat z\times {\bf s}))+ {\bf
J}\times
{\bf B}={\bf f}_{\rm el},\eqno (1)$$ and that consequently, a positive nonlocal
term
proportional to $q^2_{\perp}$ should exist in the conductivity perpendicular to
${\bf B}$. In eq. (1) (and through this paper,) ${\bf s}$ denotes the
displacement field
of vortex positions, ${\bf J}$ the external current, ${\bf B}$ the uniform flux
density
parallel to $z$ axis, $q_{\perp}$ the wavevector perpendicular to ${\bf B}$,
and
$\Gamma_1$ and $\Gamma_2$ will be given later. This equation is essentially an
extension$^{3)}$ of the single vortex dynamics to interacting vortex states.
Subsequently,
the presence of other nonlocal terms was proposed$^{2)}$ neglecting any sample
disorder,
although experiments have been performed in heavily twinned (disordered)
samples$^{4)}$, and, the presence deep in the liquid regime of a large
conductivity
nonlocal in the field direction was argued. It is important to {\it
theoretically}
clarify to what degree these proposals can be justified beyond the
phenomenology$^{1,2,5)}$ and within a basic dynamical model for type II
superconductors
such as the time-dependent Ginzburg-Landau equation (TDGL) which correctly
describes
the uniform linear dissipation in terms of Kubo formula.$^{6)}$ Actually, the
eq.(1) is
not correct in the following senses. Firstly, it is not compatible with
calculations
of conductivities based on the use of Kubo formula. Secondly, it is not clear
from eq. (1),
in which ${\bf f}_{\rm el}$ is independent of ${\bf J}$ and of the time
derivative
$\partial_t$, how the possible viscous (nonlocal) terms in the dc conductivity
are
changed as the vortex lattice is formed. Further, the dispersion $\sim
q^4_{\perp}$
of the shear mode found as a phase fluctuation formally and generally$^{7,8)}$
cannot
be seen in eq.(1). Improving this is important in order to avoid
misunderstanding
conclusions on the phase correlations in refs. 7 and 8.

In the present paper, a correct hydrodynamical approach, or equivalently, a
dynamical
harmonic
analysis on the transport in vortex states, particularly in the vortex lattice,
is
presented on the basis of TDGL. Our approach has no difficulties mentioned
above and
makes it possible to examine nonlocal conductivities above and below the
melting
transition, and several consequences of the resulting conductivities are
discussed.
As has been understood through studies$^{9)}$ on nonlocal and static linear
responses,
the use of the harmonic analysis for 3D and layered systems often leads to
misunderstandings on physics in the liquid regime, and hence, we will not
consider
situations with negligibly small positional correlations of vortices.

\smallskip
\smallskip
\leftline{\bf \S 2. TDGL hydrodynamics}

Our starting point is TDGL describing dynamics of the order parameter $\psi$ in
an
isotropic type II superconductor $$a(\gamma+{\rm
i}\gamma')\partial_t\psi+[b(|\psi|^2-
\rho_0)+a\xi_0^2(-{\rm i}\partial-{2\pi\over \phi_0} A)^2]\psi=0, \eqno(2)$$
where ${\rm curl}A=B\hat z$, $\gamma$, $a$, and $b$ are positive, $\phi_0$ is
the flux
quantum, $\xi_0$ the coherence length, and the $\rho_0$ is the spatial average
of the
mean squared order parameter. The inclusion of sample anisotropy is
straightforward
and will be done later. Since the linearized version of eq. (2) is considered
through the
paper, the noise term introducing fluctuations {\it in equilibrium} does not
have to
be included, and growths of time scales will be phenomenologically included
later as
the only fluctuation effects. Within the harmonic analysis given below,
spatially
varying gauge fields other than the external disturbance (see below) have
to be
excluded in order,  for the derivation of conductivities in vortex states,
to become
consistent with the corresponding one, for instance, of the ac conductivity
in the Meissner state
$\sigma_{ij}(\omega)=\pi\rho_{s0}\delta(\omega)\delta_{i,j}$
valid irrespective of the details of dynamics, where
$\rho_{s0}=2{({2\pi{\xi_0}}/
{\phi_0})^2}{\rho_0}a$ is the mean field superfluid density.

Particularly upon calculations of linear responses, it is convenient to work
rather
within the {\it harmonic} approximation of the corresponding Euclidean action
with gauge field
{\it disturbance} ${\delta}A$ (and with $\hbar=1$)
$$S_{\rm har}=\int_{\bf
r}\biggl[\,\beta\sum_{\Omega}\,a(\,\gamma|\Omega|+\,{\rm i}\gamma'
\Omega)|\psi_\Omega|^2+{\int_0^\beta}\! d\tau\, \biggl[\,a{\xi_0^2}\,|(-{\rm i}
\partial-{2\pi
\over \phi_0}(A+{\delta}A(\tau)))\psi(\tau)|^2$$ $$+b \biggl(
2|\psi_0|^2|\psi(\tau)|^2+
{1\over 2} (\psi_0^{*2}\psi^2(\tau)+\,{\rm c.c.}) \biggr) \biggr]\biggr].
\eqno(3)$$
Since we are interested only in frequencies and wavevectors accompanying
${\delta}A$
and summations with respect to them are not performed, calculations of
conductivities
based on the action (3) are equivalent to the linearized analysis on eq. (2)
and hence, formally
independent of temperature. Here $\tau$ is the imaginary time,
${\psi}_{\Omega}$ is
the temporal Fourier transform of $\psi(\tau)$, $\Omega$ is Matsubara
frequency,
$\beta$ the inverse temperature, and $\psi_0$ the mean field solution of
$\psi$. The
superconducting (and real) part $\sigma_{ij}$ of dc linear conductivity tensor
should
be always calculated in terms of Kubo formula $$\sigma_{ij}({\bf q})={{\partial
{\rho_{s,ij}}({\bf q},\,{\rm i}\Omega_0)}\over
{\partial{\Omega_0}}}\bigg|_{{\Omega_0}
{\to}+0}={{\partial}\over {\partial{\Omega_0}}}\,{{{\delta^2}(-\beta^{-1}{\rm
ln\,Tr}_
{\psi}\,\exp(-S_{\rm har}))}\over {{\delta}A_i({\bf
q},\Omega_0)\,{\delta}A_j(-{\bf q},
-\Omega_0)}}\bigg|_{{\Omega_0}{\to}+0}, \eqno(4)$$
where $\Omega_0$ is the external (Matsubara) frequency, $i,\, j\!=\!x$ or $y$,
and the
absence of dc uniform (${\bf q}\!=\!0$) superfluid density was used (see
below). More
generally, the ac nonlocal conductivities are defined
as the imaginary parts of the response
function ${\rho_{s,ij}}({\bf q},\,\omega+{\rm i}{0^+})$ divided by the real
frequency
$\omega$.  However, since general expressions of the ac nonlocal
conductivities are complicated and they always become the vortex
flow results in $q \to 0$ limit ( of the present formulation ),
the dc nonlocal conductivities will be only
considered below. Since the harmonic modes appearing in eq. (4) are spatial and
temporal
variations around the mean field solution linearly excited by the gauge field
disturbance, the linear response quantities in vortex states resulting from
such a
harmonic analysis$^{10)}$ are not accompanied by the temperature $\beta^{-1}$.
In this
sense, this approach is not a fluctuation theory and is invalid for the linear
dissipation parallel to $B$ which is a consequence of critical
fluctuations.$^{6)}$  On
the other hand, the vortex flow conductivities are  essentially
 mean field results.
Therefore, we focus below on the configuration perpendicular to $B$.
\vfill\eject

Before presenting general results, it is useful to first see results on the
linear
conductivities in the Landau level (LL) approach$^{7)}$
within the lowest and next-lowest LLs.
In such a high field approximation
the mean field solution and
the shear elastic mode are found within the lowest LL, and the next-lowest LL,
which
does not participate in constructing the mean field solution,$^{7,9)}$
provides the uniform displacement
of vortices (see eqs.(4.12) and (4.16) in ref.7).
Since, in constructing basis functions of eigenmodes,
Matsubara frequencies in the Euclidean action play similar roles to the
wavevectors
$q_z$ parallel to ${\bf B}$, including dynamical terms is easily performed
following ref.7 .  For later convenience, an amplitude-dominated
mode in the lowest
LL, denoted by $\delta\rho$, will also be included.
Then, to the lowest order in
$q_\perp$, the action (3) becomes $S_{LL}=S_0+S_1$, where
$$S_0 \simeq {\int_{\bf r}} \biggl[ \beta a{\sum_\Omega}(\,\gamma
{\rho_0}|\Omega||\chi_{\Omega}|^2
+\gamma'\Omega\,\delta\rho_{\Omega}^*\,\chi_{\Omega})$$  $$+{\int {d \tau}}
\biggl( {b \over 2}\delta \rho^2
+a {\rho_0}{\xi_0^2}({\partial_z}\chi-{{2\pi}\over {\phi_0}}
\delta {A_z})^2+{{C_{66}}\over 2}\,{r_B^4}({\partial_{\perp}^2}\chi)^2 \biggr
)\biggr], \eqno(5)$$  $$S_1 \simeq {{\rho_0}a \over {2r_B^2}}{\int_{\bf r}}
\biggl[ \beta{\sum_{\Omega}}
\biggl(\,\gamma|\Omega||{\bf s}_{0\Omega}|^2+\gamma'
\Omega({\bf s}^*_{0\Omega}\times{\bf s}_{0\Omega})_z \biggr)$$
$$+{\xi_0^2}{\int{d \tau}}
\biggl((\partial_z{\bf s}_
0)^2+{2 {r_B^{-2}}}({\bf s}_0+B^{-1}({\delta} {\bf A}_
{\perp 0}\times {\hat z}))^2 \biggr) \biggr], \eqno(5)'$$
where
 ${\bf s}^T$ (the transverse component of ${\bf s}$) is given$^{7,8)}$ by
${r_B^2}
({\partial_\perp}\chi \times {\hat z})$ with vortex spacing
$r_B\!=\!\sqrt{{\phi_0}/
{2\pi B}}$, ${\bf s}_0$ the uniform displacement with vanishing $q_\perp$,
$\chi$ a
longitudinal phase variable, and $C_{66}$ the resulting shear modulus.
Variational
equations with replacement ${\rm i}\Omega\to \omega$ (real frequency) give
eigenmodes
of the vortex lattice. Note that the resulting dispersion $\omega=-{\rm i}
(a\gamma{\rho_0})^{-1}{C_{66}}({q_\perp}{r_B})^4$ for the shear mode$^{7,8)}$
at zero
$q_z$ is different from that following from eq. (1), and hence that the
extrapolation$^{1,3)}$ of the uniform (or equivalently, the single-) vortex
dynamics
to the interacting case with the shear mode is not justified within TDGL.
Through the London limit (see the action (7)
given below),  the action (5) with the constraint ${\bf
s}^T\!=\!{r_B^2}({\partial_\perp}
\chi \times {\hat z})$ is found to correctly give hydrodynamical results at
longer
distances than
$$l={r_B}\sqrt{{C_{66}}{r_B^2}/2a{\rho_0}{\xi_0^2}},$$
which is typically of the order $r_B$.

The absence of the dc uniform superfluid density (helicity modulus)
is obvious from the expressions
(4) and (5)$'$,  implying that a constant twist pitch $\delta {A_0}$ is
transmuted into an
uniform displacement of vortices keeping the free energy invariant.  That is,
as also
indicated by Baym,$^{8)}$  situation is different from an elastic matter$^{1)}$
with free
energy invariant with respect to uniform displacements. Using (4) and (5)$'$
and
eliminating  ${\bf s}_0$,  we easily obtain the mean field expressions
on diagonal ($i\!=\!j$) and Hall ($i \neq j$) vortex flow conductivities
$$\sigma_{ij}({q_\perp}\!=\!0, {q_z})={\rho_{s0}}{{r_B^2}\over {2{\xi_0^2}}}
{(\gamma{\delta_{ij}}+{\gamma'}{\varepsilon_{ij}})\over
{(1+{r_B^2}{q_z^2}/2)^2}}.
\eqno(6)$$
Due to the constraint ${\bf s}^T={r_B^2}({\partial_\perp}\chi\times {\hat z})$
imposed
{\it before} introducing $\delta A$, the action (5), in which the degree of the
positional order is directly reflected, does not give any nonlocal (and
longitudinal)
conductivity even when ${q_\perp}\!\neq \!0$. It should be noticed that the
nonlocality
of the denominator in eq. (6), which results from the $|{\partial_z}s|^2$ term
of eq. (5)$'$
and is a typical example leading to a possible negative $q^2$-term$^{5)}$
in $\sigma_{ij}$, is
accompanied by a small length scale ($\sim r_B$)
insensitive to temperature and hence, is
not relevant to the ordering of the system. Consistently with eq. (6), the
nonlocal
superfluid density, namely the real part of $\rho_{s,ij}({\bf q}, \omega\!+\!
{\rm i}0^+)$, becomes $\rho_{s,ij}({q_\perp}\!=\!0, q_z)
\simeq {\rho_{s0}}\,{\delta_{ij}}\,
{r_B^2}{q_z^2}/2$ in dc limit. As pointed out in ref.9, this $q_z^2$ behavior
does
not change through a 3D melting transition.$^{11)}$

Next, let us here comment on eigen modes of the action (3) in 2D
and nondisssipative ($\gamma\!=\!0$) case, formally corresponding to the
rotating
superfluid $^4$He at zero temperature
described according to the Gross-Pitaevskii equation. Consistently with refs.12
and 13,
the order parameter in eq. (3) will be divided into the amplitude $\rho$ and
phase
$\varphi$;
$\psi=\sqrt{\rho}\exp({\rm i}\varphi)$ and, just as in the usual London
approximation,
the presence of the field-induced vortices will be taken into account through
the
topological condition$^{14)}$ on
${\partial_\mu}\varphi=({\partial_\perp}\varphi,
{\partial_\tau}\varphi)$ by neglecting any fluctuation-induced vortices. As a
result,
we obtain the following harmonic action
$$S_{\rm ph}=a{\int_{\bf r}}{\int {d \tau}} \biggl({\rm i}{\gamma'}{\delta
\rho}\,
{\partial_\tau}\chi+{\rho_0}
{\xi_0^2}({\partial_\perp}\chi-{r_B^{-2}}({\hat z} \times {\bf s}))^2+{b \over
{2a}}
\delta {\rho^2}+{{C_{66}} \over {2a}}({\partial_i}{s_j^T})^2 \biggr). \eqno
(7)$$
The only difference of this action from (5) (in 2D) is that the {\it
longitudinal}
current $${\bf v}_s\,=\,2a{\xi_0^2}({\partial_\perp}\chi\,-\,{r_B^{-2}}({\hat
z} \times
{\bf s}^T)) \eqno(8)$$
is nonzero in the action (7), which is necessary in considering longitudinal
conductivity nonlocal in directions perpendicular to the field.  When
${q_\perp}l>1$,
the minimal coupling in eq. (8) between $\chi$ and ${\bf s}^T$ is removed, and
in
principle the phase fluctuation behaves like in zero field case at such short
lengths,
while, if ${q_\perp}l<1$ and a gauge disturbance $\delta
{A_{\perp}}$(${q_\perp} \ne 0$)
is {\it not} substituted, ${\bf v}_s \simeq 0$ is established and thus, (7)
reduces to (5).
Variational equations resulting from the action (7) agree with those following
from a dual
representation used in ref.12, where a role of nonzero ${\bf v}_s$ at nonzero
frequency
was stressed in a context of superfluid $^4$He. The action (7) does not include
the term
$$S_{\rm inc}={\rm i}{\int_{\bf r}}{\int d{\tau}}{\gamma''}{{{\rho_0}a}\over
{2{r_B^2}}}({\bf s}\times
{\partial_\tau}{\bf s})_z, \eqno (9)$$
which is necessary in recovering the limit of incompressible fluid (i.e.,
$\delta \rho=0$
 and ${\bf v}_s=0$). This additional term with the coefficient $\gamma''$,
which should
become $\gamma'$,  cannot be detected in this phase-only analysis for the
action (3)
(Hereafter,
we will assume $\gamma''\!=\!\gamma'$).  The eigenvalues following from $S_{\rm
ph}\!+\!
S_{\rm inc}$ precisely coincide with those derived by Sonin$^{13)}$ and give
dispersions
${\omega^2_{\rm sh}}\simeq (a
\gamma')^{-2}\,b\,{C_{66}}({q_\perp}{r_B})^4/(1+({q_\perp}
{l_{inc}})^2)$ and $\omega_c^2 \simeq (2{\xi_0^2}/{\gamma'{r_B^2}})^2 (1+
({q_\perp}{l_{\rm inc}})^2)$ of two modes corresponding to the shear (massless)
and
compression (massive) elastic modes, respectively, where ${l_{\rm
inc}}={r_B^2}\sqrt
{b{\rho_0}/2a}/{\xi_0}$. The quadratic dispersion of $\omega_{\rm sh}$ at low
$q_\perp$
is an origin of the destruction$^{7,8)}$ of the long-ranged phase coherence in
the
vortex lattice. If ${\Gamma_1}=0$,
${\Gamma_2}=2{\pi}a{\rho_0}{\gamma''}B/{\phi_0}$,
{\it and} ${\bf v}_s=0$, the variational equation, with replacement $-{\rm
i}\tau \to t$
(real time), with respect to the shear displacement of $S_{\rm ph}+S_{\rm inc}$
becomes the transverse part of eq. (1) with no external current.
 In this case
the constraint ${\bf v}_s=0$ is different from that in the action (5) and
rather corresponds
to the limit of incompressible fluid. Following ref.13, this limit is valid
only at
shorter lengths than $l_{\rm inc}$, which is at most of the order of several
vortex
spacings in a type II superconductor near the melting line in a moderate field.

The above agreement with the well-known rotating superfluid hydrodynamics
justifies
the presence$^{7,8)}$ of the minimal-coupling in ${\bf v}_s$ between the phase
field $\chi$
and shear displacement $s^T$.  Based on this, we assume below that the ${\bf
v}_s^2$
term appearing in the London limit (7) will be found at the static level by
summing up
many higher LLs in GL harmonic analysis around the mean field state. This
should be
expected as far as mean field results in London limit are recovered in GL
theory.
When combining this picture with eqs. (5) and (5)$'$, we are naturally led to
invoking
the following action$^{15)}$ appropriate to examining nonlocal conductivities:
$$S_{\rm nl}=\beta{\sum_{\Omega}} {\int_q} \biggl[\,{\rho_0}a|\Omega|(\,\gamma
\,|\chi_
{q \Omega}|^2+{\tilde \gamma}_q{{r_B^{-2}} \over 2}|{\bf s}_{q \Omega}|^2 \,)+
{{C_{66}(q, |\Omega|)} \over 2}{q_\perp^2}|{\bf s}_{q \Omega}^T|^2 \biggr]$$
$$+a{\xi_0^2}{\rho_0}{\int_{\bf r}}{\int {d \tau}} \biggl[ \,({\partial_\perp}
\chi-{r_B^{-2}}
({\hat z}\times {\bf s}^T)\,-\,{2 \pi \over {\phi_0}}{\delta} {{\bf A}
^L_{\perp}})^2+{r_B^{-4}}({\bf s}^L+B^{-1}(\delta {\bf A}_
{\perp}^T \times {\hat z}))^2 \eqno(10)$$  $$+\,{{r_B^{-2}} \over
2}({\partial_z}
{\bf s})^2+({\partial_z}\chi-{{2 \pi} \over {\phi_0}}\delta
{A_z})^2 \biggr], $$
where $\delta {\bf A}_\perp^L$ ($\delta {\bf A}_\perp^T$) denotes the
longitudinal (transverse) part of the gauge disturbance defined {\it within the
x-y plane}. For later convenience, the shear modulus is assumed to have
possible
frequency and wavevector dependences, and the time scale of ${\bf s}$ was
phenomenologically changed taking account of a possibility that it may have
nonlocal
corrections ${\tilde \gamma}_q-\gamma$($>0$), due to an origin$^{1,2)}$ other
than the
freezing to the vortex lattice and  irrelevant  to statics of vortex states at
long distances. The amplitude mode leading to a ${\Omega^2}{\chi}^2$ term was
neglected.
The term (9) has to be included when considering a nonlocality of Hall
conductivity.
Again, the variational equation with respect to $s^T$ of the action (10) is
different
from that of eq. (1) with
${\Gamma_1}=2 \pi a{\rho_0}\gamma B/{\phi_0}$
due to the presence of nonzero ${\bf v}_s$.
Further, by neglecting the gauge disturbance $\delta A$ and examining the
dispersion
of the shear mode in the vortex lattice, it is found that, due to the presence
of the
first term $\sim |\Omega||\chi|^2$ in (10),  setting ${\bf v}_s=0$ from the
outset,
as in eq. (1),  always leads to
an erroneous result on the dispersion. Rather, at low enough $q$, the
constraint
${\bf v}_s \simeq 0$ is established as in eq. (5), and consistently, the second
(dynamical)
term $|\Omega|{|s^T|^2}$ in (10) becomes unimportant compared to the first
term.
In the action (10),  we assumed the
coefficient of $({\partial_z}s)^2$ term to be the same as that in (5)$'$.
Although this
is not correct in the usual London limit, the details of this coefficient do
not change
main results in the next section. The first term in (10) having the same form
as in the
lowest LL case (5) is required within TDGL formalism and actually  is
justified,
 because the $\chi$ variable to be related to the shear displacement at small
but
nonzero $q_\perp$ was shown$^{7)}$ to, up to the lowest order in $q_\perp$,
become a phase
change around $\psi_0$ even if higher LLs are {\it fully} included. It is clear
that
this dynamical term associated with the Goldstone (shear) mode is found only by
taking
account of interactions among vortices. On the other hand, in the derivation in
ref.3 of
the eq.(1), the origin of contributions at nonzero frequency was attributed to
variations of
the order parameter in the vicinity of each vortex core
which were assumed to be those of
a single vortex motion.
Consequently, the first (dynamical) term of the action (10) is overlooked in
such an extension of the single vortex dynamics to the vortex lattice.  As is
seen in the
next section,  due to the presence of this dynamical term,
the destruction$^{7,8)}$ of the
true off-diagonal long ranged order is reflected
even in dc conductivities under a current
perpendicular to $B$.

\smallskip
\smallskip
\leftline{\bf \S 3. Nonlocal Conductivities}

It is straightforward, using the expressions (4) and (10),
to find the nonlocal superconducting contributions to
 dc conductivities.  First, let us discuss the vortex lattice (solid) where
$C_{66}
(q\!=\!0, \Omega\!=\!0) \neq 0$.
In this case the longitudinal ($\parallel {\bf q}_\perp$) part
of the diagonal conductivity $\sigma_{xx}$ and the Hall conductivity are given
by
$$\sigma_{xx}^{(s)L}(q)=\rho_{s0}{{r_B^2} \over {2{\xi_0^2}}}\,{{{{\tilde
\gamma}_q}
{q_z^4}+\gamma{q_\perp^2}{r_B^2}{[{q_z^2}+2(l{q_\perp}/{r_B})^2]^2}/2} \over
{[{q_z^2}(1+{q^2}{r_B^2}/2)+{l^2}{q^2}{q_\perp^2}]^2}},$$
$$\sigma_{xy}^{(s)}(q)=
\rho_{s0}{{r_B^2} \over {2{\xi_0^2}}}\,{{\gamma'{q_z^2}} \over
{(1+{q_z^2}{r_B^2}/2)
[{q_z^2}(1+{q^2}{r_B^2}/2)+{q^2}{q_\perp^2}{l^2}]}}. \eqno(11)$$
using the length $l\propto{\sqrt{C_{66}}}$ defined in $\S$ 2.
In a layered material with mass
anisotropy $M/m(>1)$ and under a field perpendicular to the layers,
$q_z^2$ appearing in expressions (11)
is replaced by $2(1-{\rm cos}({q_z}s))m/M{s^2}$, where $s$ is the layer
spacing. The
transverse part $\sigma_{xx}^T$ of the diagonal conductivity merely becomes,
even in the
liquid regime, ${{\tilde \gamma}_q}\,{\sigma_{xx}}({q_\perp}\!=\!0,
q_z)/\gamma$
(using eq.(6))
,  and hence, in contrast to (11),  is not affected by the positional
correlation$^{16)}$ and
not relevant to the channel flow situation, proposed in ref.1,
where rather $\sigma_{xx}^L$
is measured.  We note that the combination $\sim {q_z^2}+{q_\perp^4}l^2$ seen
in
denominators of (11)
is the dispersion of the 3D shear (Goldstone) mode destroying the true
off-diagonal long
ranged order.$^{7,8)}$
In general, the magnitudes of the conductivities (11) significantly
depend on the $relative$
size of $|q_z|$ and $|q_\perp|$.  For instance,
when $|q_z|>l|q_\perp|/{r_B} \sim  |q_\perp|$,
both $\sigma_{xx}^{(s)L}$ and $\sigma_{xy}^{(s)}$
are well approximated by $\sigma_{ij}$ ( expressions (6))
with replacement $\gamma \to {\tilde \gamma}_q$.  On the contrary,
when $|q_z|<l{q_{\perp}^2}$,
$\sigma_{xx}^{(s)L}$ typically becomes the value at zero $q_z$ (or in 2D case);
$\gamma{\rho_{s0}}/{{q_\perp^2}{\xi_0^2}}$,
which is the same as the 2D result in zero field.$^{17)}$  Although
this result has also been suggested in ref.5, it is unclear to us whether the
authors in
ref.5 have distinguished the case $|q_z|>|q_\perp|$ from the case
$|q_z|<|q_\perp|$.
This divergent behavior is a reflection of the fact that, in this case,
the conductivity is
primarily determined by the phase variation $\chi$ and not
by the so-called
vortex motions
${\partial_t}{\bf s}={B^{-1}}({\bf E}\times {\hat z})$
implicit in the last term of the action (5)$'$,
where ${\bf E}$ is the external electric field. Correspondingly,
$\sigma_{xy}^{(s)}$
decreases like $\sim ({q_z}/{q_\perp^2})^2$ and {\it vanishes}, for arbitrary
(nonzero)
$q_\perp$, at zero $q_z$ or in 2D case.
In other words, as the nonlocality parallel to $B$
becomes negligible comparted to that perpendicular to $B$, contributions of a
finite
${\hat \gamma}_q$ are covered and overcome by the perfect positional
correlation (i.e.,
the infinite shear viscosity) making the Goldstone mode well-defined.
Particularly, these
features in $|q_\perp|>|q_z|$ are quite remarkable,  because, as mentioned in
relation to the actions (5) and (5)$'$,   their origin (i.e.,  a
nonvanishing  ${\bf v}_s$) is of higher order in the wavevector $q_\perp$ and
hence, is usually neglected.  On the other hand, these features are lost,
at fixed $q_\perp$ and $q_z$, as $l$ becomes shorter, i.e., with increasing
field. It is
consistent with the absence of the nonlocal conductivity
in the vortex lattice constructed
within the lowest LL (see a sentence following eq. (6)).
Unfortunately, we cannot predict precisely
a threshold field below which these nonlocal effects are measurable.
In addition, note that,
as seen above in $q_z \to 0$ limit, the infinite shear viscosity of the vortex
lattice
does not mean a zero value of the nonlocal resistivity.
We emphasize that the remarkable
difference between the uniform result (6) and
this large conductivity
in the case with zero $q_z$ and small ${q_\perp}(\neq 0)$ is a consequence
of the fact that, at zero $q_\perp$,
the shear displacement $s^T$ decouples with the zero mode $\chi$
and changes into the uniform
displacement.
As is seen by comparing (10) with (5)$'$,  the ac
conductivities in the vortex lattice, when ${q_\perp}\! \to \! 0$, reduces
to the vortex flow results
(6) except corrections irrelevant to positional correlations.
This is a typical example of a
crucial difference between ac and dc responses in a kind of ordered state.

The above results that ${\sigma_{xx}^{(s)L}} \sim {q_{\perp}^{-2}}$ and
${\sigma_{xy}^{(s)}}=0$ when ${q_z}=0$ and ${q_{\perp}} \neq 0$ are valid even
for a
state with nonzero ${C_{66}}({q_\perp}\! \neq \! 0, |\Omega|\!=\!0)$ such as
the 2D
hexatic liquid state which may be possible in low enough fields (The 3D hexatic
phase
invoked in refs.1 and 10 is inconsistent with the first order freezing to the
vortex
lattice and never realized$^{18)}$ in real superconductors, and hence need not
to be considered). They are also applicable to real systems with pinnings and
hence
with a finite correlation length $\xi_h$ of the bond orientational order,
if ${q_\perp}{\xi_h}>1$. Therefore, it is valuable to examine these dc linear
responses in real experiments in relation to determinations of the position
of the freezing to the vortex lattice and of the existence or absence
of the hexatic phase in 2D systems. In particular, it should be noted that the
behavior
${\sigma_{xx}^{(s)L}} \sim {q_{\perp}^{-2}}$ at nonzero $q_\perp$ as well as
the
vanishing $\sigma_{xy}^{(s)}$ implies ${\rho_{xy}} \simeq
{\rho_{xx}^2}\,{\sigma_{xy,N}}$, where ${\rho_{xx}}({\rho_{xy}})$ is the total
diagonal (Hall) nonlocal resistivity, and $\sigma_{xy,N}$ is the extrapolated
normal
Hall conductivity.

When trying to understand the liquid regime (disordered phase) in the present
approach,
 some comments are necessary.  As shown in ref.9, the harmonic analysis cannot
be
used in the situation where the contributions with nonzero reciprocal lattice
vectors
are negligible even if the amplitude fluctuation of $\psi$ around $\psi_0$ is
negligible. Even in such situations the nonvanishing transverse diamagnetic
susceptibility results from the static superconducting fluctuation$^{6)}$ and,
in layered
systems,  shows a dimensional crossover$^{9)}$ due to a competition between the
layer
spacing and a finite phase coherence length parallel to $B$ which cannot be
detected
in the harmonic analysis. Consistently, this dimensional crossover will be seen
also
in the nonlocal conductivity. Unfortunately, it is practically difficult at
present to
find possible nonlocal corrections,  associated with  the freezing
to the vortex lattice,  to the vortex flow expression (6) according to the
fluctuation
theory,$^{6)}$ but the mean field (i.e., harmonic) analysis may become a
guideline to an
extension of the analysis in ref.6.  In addition, it is possible that the
present
analysis based on the phase-only model (i.e., based on the mean field solution)
will
provide an essentially correct result at relatively {\it short} scales in
directions
perpendicular to $B$ in the disordered state.$^{9)}$  For these reasons, we
will focus on
the region just above the freezing point to the vortex lattice and only
consider
the nonlocal conductivities with vanishing $q_z$ (see, however, $\S$ 4),
in which case the dimensional crossover does not appear, and hence, the use of
the
harmonic approximation may be permitted.

In this case, the corresponding results to (11) are given by
$$\sigma_{xx}^{(l)L}({q_\perp}; {q_z}=0)={\rho_{s0}}{{r_B^2}\over
{2{\xi_0^2}}}\,{{{{\tilde \gamma}_q}+
\eta{q_\perp^2}{r_B^2}/{{\rho_0}a}}\over {1+{q_\perp^2}{r_B^2}({{\tilde
\gamma}_q}+
\eta{q_\perp^2}{r_B^2}/{{\rho_0}a})/{2 \gamma}}},$$
$$\sigma_{xy}^{(l)}({q_\perp}; {q_z}=0)=\rho_{s0}
{{r_B^2}\over {2{\xi_0^2}}}\,{{\gamma'}\over {1+{q_\perp^2}{r_B^2}({{\tilde
\gamma}_q}+\eta{q_\perp^2}
{r_B^2}/{{\rho_0}a})/{2 \gamma}}}, \eqno(12)$$
where the Maxwell form$^{19)}$ $C_{66}(q, |\Omega|) \simeq \eta|\Omega|$ with
shear
viscosity $\eta$ was
assumed$^{20)}$ for the dynamical shear modulus. Interestingly, the expressions
(12) {\it smoothly} lead
to the results (11) for the vortex lattice  at zero $q_z$  by taking $\eta$ to
be infinity.
The terms $1+\eta({q_\perp}
{r_B})^4/2\gamma {\rho_0}a$ in the denominators of expressions (12) again
originates,
through a nonzero ${\bf v}_s$ (i.e., a {\it higher} spatial gradient),
from the spectrum of the shear mode $\omega \sim
-{\rm i}{C_{66}}({q_\perp}{r_B})^4$, which cannot be detected in eq. (1),
 while the nonlocality in the
numerator of $\sigma_{xx}^{(l)L}$,
resulting from nonlocalities of the time scale for the displacement
field, is the same as that expected from (1).  As a result, at large $q_\perp$
of the order $r_B^{-1}$, both (11) and (12) are well approximated by the usual
one (6),
suggesting that, for such
a rapid variation of the external current, not only collective effects but also
the
sample disorder
(pinning) is irrelevant as far as the time scale to be affected by the sample
disorder is
${\tilde \gamma}_q$ and not $\gamma$.  In general, in the case with a pinning
effect
enhancing ${\tilde \gamma}_q$, the viscous effect accompanying the freezing to
the
vortex lattice
becomes negligible.$^{1)}$  Further, the expressions (12) suggest that the low
temperature limits of the
nonlocal conductivities with nonzero $q_\perp$ in 2D systems with pinnings,
where we have no transitions
at nonzero temperatures,  again become the corresponding, and above-mentioned,
results (independent of
${\tilde \gamma}_q$ ) in the vortex lattice, again leading to the relation
$\rho_{xy}
\simeq {\rho_{xx}^2}{\sigma_{xy,N}}$.
This relation should also be observed in a clean 3D-like situation
with vanishing $q_z$ and a large $\eta$. An apparently similar relation was
found$^{21)}$ in the
pinning-dominated region of BSCCO.

\smallskip
\smallskip
\leftline{\bf \S 4. Discussions and Conclusion}

We will discuss consequences of possible viscous effects$^{1,2)}$ deep in
the liquid regime of pinning
free systems, {\it other than} $\eta$,  which should appear through
nonlocalities
of ${\tilde \gamma}_q$.
According to eqs. (11) and (12), when $|q_z| \ll |q_\perp|$, it is difficult to
practically divide, in
$\sigma_{xx}^{(l)L}$,  contributions of ${\tilde \gamma}_q$ from the positional
correlation effects,
and hence,  we will only consider the case with vanishing $q_\perp$ but nonzero
$q_z$,
where the diagonal
conductivities can be always expressed by (6) with replacement $\gamma \to
{\tilde \gamma}_q$. Further
we will focus on the 3D region below the dimensional crossover$^{9)}$ (We note
that this dimensional
crossover, formally equivalent to that resulting from the entanglement
picture,$^{1)}$ has nothing to
do with that argued through experiments in ref.4). For this case, it is often
argued$^{22)}$ that
thermally activated cutting (and reconnection) processes among vortices induced
by a nonuniform
(${q_z} \neq 0$) current will lead to a very long length scale and
significantly increase
${\tilde \gamma}_q$ even if the (thermally-induced) entanglement is
absent.$^{2)}$
This picture seems to suggest
that the time scale grows unlimitedly even in the vortex lattice upon cooling,
and hence that the
vortices bent by the current cannot move at low enough temperatures. It should
be noted that such an
argument would also be applicable
to a nondissipative case (with zero $\gamma$ but nonzero $\gamma'$) by
imagining a nonlocal growth of $\gamma'$. However, it is even unclear to us if
this is a correct
argument in the context of {\it linear} hydrodynamics. In superfluid $^4$He,
the reconnection
process does not$^{23)}$ seem to need a remarkable slow dynamics. In addition,
the algebraic
system-size dependence of an onset temperature$^{4,24)}$
of apparently nonlocal vortex motions is inconsistent with
the argument based on the thermally activated vortex cutting processes
in which the temperature
variation is dominated by the Arrenius factor, because the measured algebraic
size dependence
inevitably means an algebraic decrease of the cutting barrier itself with
increasing the system-size,
although intuitively such a decrease of the barrier should not be expected.
Further,
the resistivity data$^{24)}$ parallel to $B$ in a heavily twinned sample seem
to intuitively
contradict those in a more 3D-like situation.$^{25)}$  Clearly, experiments
in twin-free samples are necessary.$^{26)}$
Theoretically, there seems to be no acceptable reasons why such a nonlocal
growth
of ${\tilde \gamma}_q$ has to be expected.  Firstly,  our
analysis shows that, in contrast to the phenomenology in ref.2 and 5,  viscous
terms
possible in
${\tilde \gamma}_q$ cannot reduce to elastic terms in the vortex lattice.
Secondly, the (if any) entanglement$^1$ making a growth of ${\tilde \gamma}_q$
possible must disappear at least deep in the
vortex lattice.$^{27)}$  In ref.9,
the presence of a large and positive nonlocal ($\sim {q_z^2}$)
contribution in ${\tilde \gamma}_q$ was questioned on the basis of the
observation that,
in contrast
to the pinning-induced activation form for the time scale, any nonlocal (i.e.,
$q_z$-dependent)
growth of the time scale may be incompatible with the uniform vortex flow due
to the nonlinearity
(i.e., summations with respect to internal $q_z$'s) in Kubo formula in the
nonGaussian fluctuation
theory.$^{6)}$  At the present stage, however, this opinion is not still
conclusive.

As already commented in the context,
the present analysis cannot clarify a characteristic field
below which the positional
correlation-induced nonlocal responses studied in this paper and the resulting
relation $\rho_{xy}
\sim {\rho_{xx}^2}$ become measurable.  In order to estimate
this theoretically,  it is necessary to know how a nonzero ${\bf v}_s$,
required through the agreement with the
rotating superfluid hydrodynamics,  is found consistently with the action (9).
{}From the fluctuation
theory,$^{6)}$  the nonlocality in the numerator of $\sigma_{xx}^{(l)L}$ may be
expected to be detected as
a consequence of renormalizations of the time scale for vortex motions. On the
other hand, it is
extremely difficult to detect nonlocalities in the denominators of
conductivities (12), because
their origin consists in the details of physics at short scales which, as
mentioned in ref.9, cannot
easily be captured in the fluctuation theory from higher temperatures.

In conclusion, a hydrodynamics for vortex states in type II superconductors
consistent with the
rotating superfluid hydrodynamics has been presented in order to study the
nonlocal conductivities
perpendicular to the magnetic field.  In particular, the conductivities with
nonlocality only in
perpendicular directions to the field and in a situation with large shear
viscosity are essentially
the same as those in zero field case, and the vanishing superconducting part of
the Hall conductivity
under such a situation will be useful in experimentally judging the existence
or absence of the 2D
hexatic phase in type II superconductors.  It is interesting to experimentally
examine the predicted
relation $\rho_{xy} \simeq {\rho_{xx}^2}{\sigma_{xy,N}}$ not only in dirty
samples but in a clean
sample such as untwinned YBCO.

\vfill\eject
\leftline{\bf Acknowledgement}

The author acknowledges related discussions with Alan Dorsey, Wai Kwok, and
Ulrich Welp, and is
grateful to Physics Department, Indiana University for hospitality where this
manuscript was written.
This research was finantially supported by Grant-in-Aid for Scientific Research
from the Ministry of
Education, Science, and Culture in Japan.

\bigskip
\bigskip
\bigskip

\leftline{\bf References}
\frenchspacing

\item{1)} M. C. Marchetti and D. R. Nelson: Phys.Rev.B {\bf 42}(1990) 9938.
\item{2)} D. A. Huse and S. N. Majumdar: Phys.Rev.Lett. {\bf 71}(1993) 2473.
\item{3)} See, for instance, L. P. Gor'kov and N. B. Kopnin: JETP {\bf
33}(1971) 1251.
\item{4)} H. Safar et al.: Phys.Rev.Lett. {\bf 72}(1994) 1272.
\item{5)} C. -Y. Mou, R. Wortis, A. T. Dorsey, and D. A. Huse: preprint.
\item{6)} R. Ikeda, T. Ohmi, and T. Tsuneto: J.Phys.Soc.Jpn. {\bf 60} (1991)
1051.
\item{7)} R. Ikeda, T. Ohmi, and T. Tsuneto: J.Phys.Soc.Jpn. {\bf 61} (1992)
254, and
ibid {\bf 59} (1990) 1740.
\item{8)} M. A. Moore: Phys.Rev.B {\bf 45} (1992) 7336; G. Baym: preprint.
\item{9)} R. Ikeda: J.Phys.Soc.Jpn. {\bf 64} (1995) 1683.
\item{10)} T.Chen and S.Teitel: Phys.Rev.Lett. {\bf 72} (1994) 2085.  It is
clear that,
as done by these authors, imposing the London gauge to the gauge {\it
disturbance}
in calculations of linear responses is misleading and has no theoretical
foundation,
because it would consistently require the absence of any longitudinal component
($\sigma_{xx}^L$) of the dc conductivity which has to exist in order to become
consistent
with the uniform (${\bf q}=0$) case (see also a sentence following eqs.(11)).
The
absence of the longitudinal component of the superfluid density tensor at
{\it zero frequency} is easily found without any restriction on the gauge
disturbance.
\item{11)} This means that the elastic terms associated with the compression
mode are not affected
by the 3D melting transition (see ref.9). It also invalidates a corresponding
result $\rho_{s,xx}
({q_\perp}\!=\!0, q_z) \sim |q_z|$ [M. V. Feigel'man and L. B. Ioffe: JETP
Lett.
{\bf 61} (1995) 75] in a putative liquid phase resulting from the boson analogy
even if
the fluctuating magnetic field is neglected,  because the divergent
${\chi_\perp^{(c)}}\,\sim {\rho_{s,xx}}({q_\perp}\!=\!0, q_z)/{q_z^2}
\sim|q_z|^{-1}$
resulting from it would mean that such an intermediate phase has a stronger
ordering than
 the {\it unpinned} vortex lattice in which $\chi_\perp^{(c)}$ is finite (see
refs.9 and 16).
Namely, the intermediate liquid phase with no linear dissipation parallel to
$B$ is compatible not with the {\it unpinned} vortex lattice but with a {\it
pinned} vortex lattice
with nonzero $\rho_s({\bf q}\!=\!0)$ (i.e.,
${\chi_\perp^{(c)}}\!\sim{q_z^{-2}}$).  This
is quite consistent with the simulation results of Teitel and coworkers[Phys.
Rev. B {\bf 49}
 (1994) 4136], where the unpinned 3D vortex lattice was not found.
\item{12)} S.C.Zhang: Phys.Rev.Lett. {\bf 71} (1993) 2142.
\item{13)} See eqs. (4.71) to (4.73) in E. B. Sonin: Rev.Mod.Phys. {\bf 59}
(1987) 87.
Parameters $c_T$, $c$, and $\Omega$ appearing in this reference are given using
our
notation by $c_T^2 \equiv 2{\xi_0^2} {C_{66}}/a{\rho_0}{\gamma'^2}$, $\Omega
\equiv {\xi_0^2}/{{r_B^2}\gamma'}$, and $c^2 \equiv 2b{\rho_0}{\xi_0^2}
/a{\gamma'^2}$.
\item{14)} See Appendix D in ref.7,  and V.N.Popov: {\sl Functional Integrals
and Collective Excitations } (Cambridge, UK, 1987), Ch.8.
\item{15)} See Appendix D in ref.7,  and  A. I. Larkin and Y. N. Ovchinikov:
J.Low Temp.Phys. {\bf 34} (1979) 409.
\item{16)} Correspondingly, we have two kinds of the transverse parts of
diamagnetic susceptibility (, closely related to the tilt moduli as mentioned
in Ref.9,)
due to the presence of two (elastic)
modes, and they are given by
$\chi_\perp^{(s)}\!=\!{r_B^2}\,{\rho_{s0}}({q_z^2}+2(l{q_\perp}/{r_B})^2)/
({q_z^2}(2+(q{r_B})^2)+2{l^2}{q_\perp^4})$ and $\chi_\perp^{(c)} \simeq
\!{r_B^2}{\rho_{s0}}/2$ for
the shear and compression modes, respectively. Note that the divergent behavior
of $\chi_\perp^{(s)}(
{q_\perp} \to 0; {q_z}=0)$ does not contradict the vanishing superfluid density
perpendicular to $B$.
Since we have neglected in eqs. (5)$'$ and (10) an O(${q_\perp^2}{s^{L2}}$)
term playing no significant
roles in nonlocal linear dissipations, the longitudinal susceptibility is zero
in the present approach.
\item{17)} H. J. Mikeska and H. Schmidt: J.Low Temp.Phys. {\bf 2} (1970) 371.
\item{18)} E. Frey, D. R. Nelson, and D. S. Fisher: Phys.Rev.B {\bf 49} (1994)
9723.
\item{19)} L. D. Landau and E. M. Lifshitz: {\sl Theory of Elasticity, 3rd
Edition}
(Pergamon, Oxford, 1986), sec.36.
\item{20)} It may be possible that, as far as the wavevector $q_z$ is nonzero,
the static
(zero frequency) shear modulus deep in the liquid regime is nonzero and has the
form $C_{66}\!\simeq\!
{q_z^2}{r_B^2}f(q)$, where $f(0)=0$. Note a similarity between $|\Omega|$ and
$q_z^2$ in the action
(10). In this case the susceptibility $\chi_\perp^{(s)}$ (see Ref.16) is
slightly enhanced when
${q_\perp} \neq 0$: ${\chi_\perp^{(s)}}({q_\perp} \neq 0) \simeq
{\rho_{s0}}{r_B^2}(1+
2{q_\perp^2}{r_B^2}f(q))/(2+2{q_\perp^4}{r_B^4}f(q))$.
\item{21)} A. V. Samoilov: Phys.Rev.Lett. {\bf 71} (1993) 617.
\item{22)} For instance, C. Carrano and D. S. Fisher: Phys.Rev.B {\bf 51}
(1995) 534.
\item{23)} K.W.Schwartz: Phys.Rev.B {\bf 38} (1988) 2398.
\item{24)} D.Lopez et al.: Phys.Rev.B {\bf 50} (1994) 7219.
\item{25)} W.K.Kwok et al: Phys.Rev.Lett. {\bf 64} (1990) 966.
\item{26)} According to a preliminary measurement [F. de la Cruz: in {\sl Int.
Workshop on
Vortex Dynamics, Lake Forest, June 1995} (unpublished)], the nonlocal effect on
the
resistivity just above the melting transition in an {\it untwinned} YBCO sample
is not seen.  This is consistent with ref.11 and with the
statement in ref.4 that the nonlocal effect has something to do with
the `shoulder' in a resistivity curve due to the twin-boundary pinning.
\item{27)} In addition, the picture in ref.1 that the thermally-induced
entanglement will become more
remarkable with increasing field is questionable, because a situation dominated
by the lowest LL mode,
in which the entanglement is absent (see ref.9 and M. J. W. Dodgson and M. A.
Moore:
preprint),  must become valid with increasing field.

\bye